\documentclass[aip,apl,floatfix,reprint]{revtex4-1}
\usepackage{graphics}
\usepackage{bm}
\usepackage{graphicx}

\begin{document}
\title{Coupling of silicon-vacancy centers to a single crystal diamond cavity} 
\author{Jonathan C. Lee}\
\author{Igor Aharonovich}
\author{Andrew P. Magyar}
\author{Fabian Rol }
\author{Evelyn L. Hu}
\email{ehu@seas.harvard.edu}
\affiliation{School of Engineering and Applied Sciences, Harvard University, Cambridge, Massachusetts, 02138}
\date{\today}

\begin{abstract}
Optical coupling of an ensemble of silicon-vacancy (SiV) centers to single-crystal diamond microdisk cavities is demonstrated. The cavities are fabricated from a single-crystal diamond membrane generated by ion implantation and, electrochemical liftoff followed by homo-epitaxial overgrowth. Whispering gallery modes which spectrally overlap with the zero-phonon line (ZPL) of the SiV centers and exhibit quality factors $ \sim $ 2200 are measured. Lifetime reduction from 1.8 ns to 1.48 ns is observed from SiV centers in the cavity compared to those in the membrane outside the cavity. These results are pivotal in developing diamond integrated photonics networks.
\end{abstract}

\maketitle

Color centers in diamond have emerged as leading candidates for solid state quantum information processing (QIP) due to their unique physical properties such as single photon emission at room temperature\cite{kurtsiefer_stable_2000, neumann_single-shot_2010, buckley_spin-light_2010, togan_quantum_2010}. Intense research efforts on individual color centers in diamond over the past decade have yielded significant insights into the photo-physics of defects in diamond. In particular, there is a growing interest in utilizing color centers that exhibit a narrow luminescence band, short excited state lifetime, and single photon emission at room temperature\cite{neu_single_2011, aharononvich_chromium_2010, rabeau_fabrication_2005}. Such defects are extremely attractive photon sources for quantum key distribution or optical quantum computation\cite{gisin_quantum_2002}. One example of such a defect is the SiV center, consisting of a single silicon atom in the split vacancy configuration with a ZPL transition at 738 nm\cite{goss_twelve-line_1996} as shown in Fig. 1a. 

Significant emphasis has been devoted to the coupling of emitters to optical cavities to facilitate the control of emission properties while guiding light into a photonic network\cite{aharonovich_diamond_2011, santori_nanophotonics_2010, faraon_resonant_2011, barclay_hybrid_2011, babinec_diamond_2010, wolters_enhancement_2010, englund_deterministic_2010, van_der_sar_deterministic_2011}. However, a major challenge to achieving diamond-based quantum photonics is the difficulty in processing single-crystal diamond for the fabrication of optical cavities coupled to its color centers. Indeed, fabricating such cavities from bulk single crystal diamond by ion implantation degrades the optical properties of the embedded color centers impeding their use for photonic platform\cite{fairchild_fabrication_2008, bayn_triangular_2011}. Furthermore, earlier work has shown that using nanocrystalline diamond to form optical cavities places limits on the quality factors of those cavities\cite{ wang_fabrication_2007}. Even for epitaxial material systems where very high quality factors ($ > $ 10$^5$) have been demonstrated, ‘subtle’ variations in materials quality and absorption are found to still limit the Q obtainable\cite{michael_wavelength-_2007}. Thus, it is critical to form high quality optical cavities from single crystal diamond, but this in turn poses additional challenges. 

This work addresses some of those challenges through the formation of high quality, optically thin, single crystal diamond films incorporating SiV centers formed by epitaxial overgrowth on a single crystal membrane. Microdisk cavities were subsequently fabricated from the material, demonstrating quality factors in the range of 2000-3000. Cavity-mediated reduction of the fluorescence lifetime of the SiV centers was also measured. 

The starting material was a type II-a CVD diamond sample, obtained from Element-Six. Diamond membranes, 1.7 $\mu$m thick, were generated using ion implantation followed by thermal annealing and an electrochemical etch process\cite{magyar_fabrication_2011}. The membranes were cleaned using a mixed (1:1:1 sulfuric:perchloric:nitric acid) boiling acid bath, and then transferred onto a 2 $\mu$m thick silicon dioxide layer on top of silicon, which provides optical isolation and allows independent optical characterization of the diamond membranes. Photoluminescence (PL) from the diamond membranes showed no optical signatures of the SiV centers. A short epitaxial over-growth ($\sim$ 200 nm) was employed using a  microwave plasma chemical vapor deposition (MPCVD) reactor to introduce the SiV centers\cite{aharonovich_homoepitaxial_2011}, the surface roughness is 3 nm after the regrowth. The overgrown material is grown epitaxially, and forms a complete single crystal. Neither polycrystalline material nor grain boundaries have been observed in our membranes. The composite structure was flipped so that the original template is the top surface. The structure was thinned using argon-chlorine-based inductively coupled plasma reactive ion etching (ICP RIE) to first smooth the surface\cite{lee_etching_2008}, followed by an oxygen-based ICP RIE to thin the structure to $\sim$ 500 nm thick. Thus, the resulting structure is a composite diamond membrane composed of 300 nm of the original membrane template and 200 nm of the overgrown material. The schematic of the fabrication process is depicted in Fig. 1c.

\begin{figure}[b]
 \includegraphics[width=8.5cm]{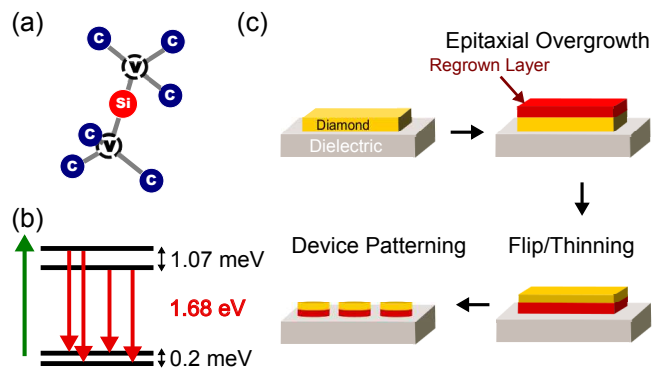}%

 \caption{ (color online) (a) Schematic of the SiV defect structure in diamond. The red atom represents the silicon, the empty circle represents the surrounding vacancies and the blue atoms are the carbon atoms. (b) The energy level diagram of SiV centers with doublets in both the ground states and excited states. The emitter can be excited by a 532 nm excitation and the emission is centered near the ZPL at 738 nm. (c) Fabrication process of the cavity. A $\sim$ 1.7 $\mu$m thick diamond membrane is placed on a silicon-dioxide substrate followed by epitaxial overgrowth of a thin diamond film. The film is flipped and thinned using reactive ion etching and the microdisk cavities are patterned using e-beam lithography with silicon-dioxide as a hard mask.}%
 \end{figure}

Reduced Raman linewidth was measured from the overgrown layer (full width half maximum (FWHM):  $\sim$ $3cm^{-1}$) compared to the original diamond template (FWHM:  $\sim$ $9 cm^{-1}$) indicating the better material quality of the overgrown layer. Fig. 2a shows the room temperature PL spectra of the diamond membrane before thinning (blue curve) and after the removal of $\sim$ 1.4 $\mu$m (green curve) and $\sim$ 1.6 $\mu$m (red curve) of the original diamond template. Broadband luminescence obscured the luminescence of SiV centers before the membrane is thinned (Fig. 2a blue curve). The broadband luminescence is further reduced through the removal of the original diamond template (Fig. 2a green and red curves). A pronounced ZPL at 738 nm attributed to the SiV is clearly seen after most of the original template was etched away. The FWHM of the ZPL was measured to be $\sim$ 6 nm at room temperature. At low temperature (18 K) the FWHM was reduced to $\sim$ 3 nm (Fig. 2b), however, the fine structure of the defect (fig. 1b) was not resolved\cite{sternschulte_1.681-ev_1994, clark_silicon_1995} due to a high concentration of the emitters and their inhomogeneous broadening due to the strain fields.

Microdisk cavities were made from the 500 nm thick diamond membrane after epitaxial overgrowth and ICP-RIE thinning. An 80 nm thick silicon dioxide hard mask was deposited using plasma-enhanced chemical vapor deposition (PECVD). The 2.5 $\mu$m diameter microdisks were patterned using a 100 KeV electron beam lithography system (Elionix) with poly (methyl methacrylate) (PMMA) as the resist. The patterns were then transferred by a fluorine based ICP-RIE step to etch into the silicon dioxide hard mask, followed by a oxygen based ICP-RIE step to transfer the pattern to the diamond membrane. Inset of fig. 2c shows a scanning electron microscope (SEM) image of the diamond microdisk cavity with a diameter of 2.5 $\mu$m and a thickness of 500 nm. Fig. 2c shows the room temperature micro-PL measurements recorded from the 2.5 $\mu$m diameter disk. The PL measurements were performed with a confocal microscope (LabRAM ARAMIS, Horiba Jobin-Yvon) with 532 nm laser excitation. The excitation and signal collection went through the same objective with NA = 0.9 and magnification 100$\times$ which results in spatial resolution $\sim$ 1 $\mu$m. By comparing the free spectral range of the whispering gallery modes (WGMs) to the Finite difference Time Domain (FDTD) simulations, the polarization and the order (both radial and azimuthal direction) of the WGMs are identified. Quality factors, Q, of these modes were determined to be $\sim$ 2200 by calculating $\frac{\lambda_{cav}}{\Delta\lambda_{cav}} $, where $ \lambda_{cav} $ is the cavity mode wavelength and $\Delta\lambda_{cav}$ is the FWHM of the mode.

\begin{figure}[b]
 \includegraphics[width=8.5cm]{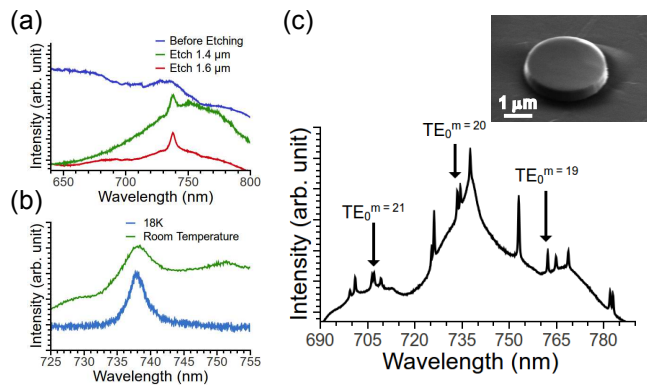}%

 \caption{ (color online) (a) Room temperature PL curves recorded from the membrane immediately after regrowth (blue curve), after a subsequent removal of $\sim$ 1.4 $\mu$m of the original template (green curve) and a final thinning to 300 nm thick membrane (100nm:200nm original template:overgrown layer). A pronounced ZPL from the SiV centers was observed as the original template is removed. The FWHM of the ZPL is $\sim$ 6 nm. (b) Low temperature PL measurement recorded from the membrane after removal of $\sim$ 1.4 $\mu$m of the original membrane. The FWHM of the ZPL of SiV centers at 9 K is $\sim$ 3 nm.  (c)  SEM image of a 2.5 $\mu$m diameter micro-disk cavity fabricated from single-crystal diamond. Room temperature PL spectrum recorded from the microdisk cavity. The WGMs are observed. The 1st order radial TE mode has a Q of $\sim$ 2000. }%
 \end{figure}

A high resolution spectrum of the TE mode centered at 736.5 nm is shown in Fig 3a. The mode is found to be zeroth order in the radial direction (TE$_0^m=20$) with Q $\sim$ 2200 and overlaps spectrally with the SiV ZPL emission. To investigate the emitter-cavity system further, lifetime measurements were performed. A frequency-doubled pulsed Ti:Sapphire laser at 460 nm with a 76 MHz repetition rate and pulse width less than 70 femtosecond was used to excite the SiV emitters. The 18K measurement was made with the sample in a cryostat (Janis) with excitation and signal collection passing through the same NA = 0.5 objective, with 100$\times$. The reflected laser light was reflected by a dichroic mirror and the collected, PL signal was filtered by a band pass filter (747 nm $\pm$ 17 nm) and directed onto an avalanche photo-diode (Micro Photonic Devices, jitter time $\sim$ 50 ps). The results of the lifetime measurements are shown in Fig. 3b and are best fit with a bi-exponential decay. The lifetime of the SiV centers coupled to the microdisk is measured to be $\sim$ 1.48 $\pm$ 0.04 ns ns while the lifetime measured from the membrane outside of the disk is $\sim$ 1.83 $\pm$ 0.09 ns. The second fast exponential decay accounts for the fast decays which is measured 0.66 $\pm$ 0.02 ns from the membrane, and 0.43 $\pm$ 0.01 ns from the cavity due to other defects in the membranes. The lifetime reduction is in reference to a diamond membrane having the same thickness and having undergone exactly the same fabrication process as the microdisk cavities. Hence, other reasons for lifetime reduction, e.g. non radiative channels or surface defects, if present, would be the same in both the membrane and the microdisk. Therefore, we believe that the only reason for a lifetime reduction is the modification of the density of states of the emitters due to coupling to the cavity modes. The Purcell enhancement based on lifetime modification is estimated to be $\sim$ 1.3. The reduced lifetime value can be regarded as an averaged lifetime of multiple different exponential decay curves, which represent different SiV centers in the disk coupled to the cavity mode. The photon count rate from the avalanche photo diode (APD) was $\sim$ 40,000 counts/s and 3000 counts/s when collected from the microdisk and the diamond membrane, respectively, under identical excitation conditions. The increased emission can be attributed to increased collection efficiency due to scattered light from the microdisk cavity. However, the reduced lifetime of the SiV centers within the cavity is evidence of the coupling of the ensemble of the emitters to the microdisk cavity. 

To estimate a reasonable range of lifetimes for SiV centers in the cavity, given that the Q of the mode is $\sim$ 2200, we consider the spontaneous emission rate enhancement of an emitter inside a cavity, described by the Purcell factor: $ F_p = \frac{3}{4\pi^2} (\frac{\lambda}{n})^3 \frac{Q}{V} $. For this analysis, we have neglected non-radiative decays, which may be present\cite{turukhin_picosecond_1996}. To achieve optimal coupling between the cavity and dipole, the emission dipole needs to spatially overlap with the maximum field and have the same orientation as the electric field of the cavity mode. The modal volume of the TE$_0^{m=20}$ is $\sim$ 0.273 $\mu$m$^3$ based on FDTD simulation. The measured linewidth of the mode is $\sim$ 0.336 nm which means that a maximum of $\sim$10 $\%$ of the SiV ensemble may overlap spectrally with the cavity. This assumption is reasonable due to the large Debye-Waller factor of the SiV center, which results in $\sim$ 80$\%$ of the emission concentrated in the ZPL\cite{neu_single_2011}. Under these conditions, the maximum Purcell enhancement for a single emitter is estimated to be F $ \sim $ 17. The cavity mode is coupled to an ensemble of emitters where the spectral overlap, dipole orientations and spatial overlap could varied. Therefore, we believe that the $ \sim $ 1.3 reduction in spontaneous emission lifetime observed for the cavity-coupled SiV centers is reasonable for our system.

\begin{figure}[b]
 \includegraphics[width=8.5cm]{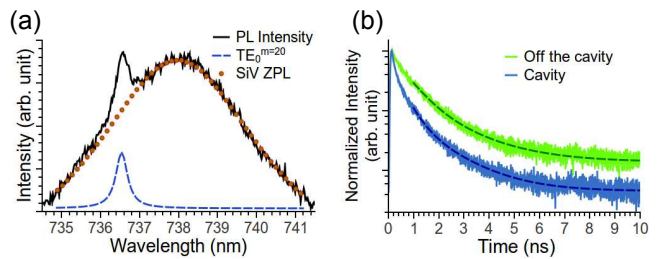}%

 \caption{ (color online) (a) Room temperature PL spectrum of the 1$^{st}$ order radial TE mode (TE$_0^m=20$). The Q factor is $\sim$ 2200 and spectral overlap with the ZPL of SiV centers was observed. (b) Fluorescence lifetime measurement of SiV centers within microdisk resonators shows lifetime reduction from 1.83 ns to 1.48 ns, compared to the SiV centers in the diamond membrane, recorded at 18 K. The lifetime was fit to a bi-exponential decay model (dash lines) where reduction for both the fast and slow decay channel were observed. }%
 \end{figure}

The quality factors of $\sim$ 2200 measured from the microdisks are comparable with values recently reported for a single crystal diamond micro-ring resonator\cite{faraon_resonant_2011} and a significantly higher value than devices made of nanocrystalline diamond\cite{wang_fabrication_2007} or milled by using a focused ion beam\cite{bayn_triangular_2011}. Examination of the PL data of Fig. 2a, shows the SiV peak sitting atop a broad band of luminescence, which is successively diminished as the composite diamond membrane is thinned, the SiV peak sitting atop a broad band of luminescence. The broad band luminescence is likely due to other impurities within the membrane and could lead to material reabsorption and hence limit the Q of the cavity. Removal of that central region may also provide an increase in the observed Q as has been seen in other material systems\cite{tamboli_observation_2009}.

Coupling of SiV defects to an optical cavity marks pivotal progress towards the realization of scalable diamond-based quantum photonics networks. Although demonstrated with an ensemble in this work, coupling of single emitters should also be possible using the techniques described here. Our approach enables the formation of optically thin single crystal membranes of good optical and structural properties. Such membranes can serve as the basis for microdisks and photonic crystal cavities as well as enable the construction of an integrated photonic network of coupled diamond cavities and waveguides. The approach provides flexibility in both the formation of color centers, as well as flexibility in the geometry of optical cavities formed around those centers. We believe that further optimization of the fabrication steps, such as the use of high temperature annealing or further reduction of impurity absorption will produce high Q’s and stronger emitter-cavity coupling. These initial results represent an important milestone in the achievement of diamond-based cavity quantum electrodynamics. 

The authors acknowledge the help of D.R. Clarke for access to the PL and Raman facilities, and M. Huang for assistance with ion implantation. The authors also thank K.J. Russell, T.L. Liu, T.M. Babinec, A.L Falk, H.J. Heremans, B.B. Buckley, C.G. Yale and D.D. Awschalom for useful discussions. This work was carried out with the financial support of DARPA under the Quantum Entanglement Science and Technology (QuEST) Program. This work was enabled by facilities available at the Center for Nanoscale Systems (CNS), a member of the National Nanotechnology Infrastructure Network (NNIN), which is supported by the National Science Foundation under NSF award no. ECS-0335765. CNS is part of the Faculty of Arts and Sciences at Harvard University. At the time of manuscript preparation a related preprint appeared, which describes coupling of SiV to photonic crystal cavities\cite{riedrich-moller_one-_2011}.

\end{document}